Title: Successful Recovery of an Observed Meteorite Fall Using Drones and Machine Learning.

Seamus L. Anderson[1*], Martin C. Towner[1], John Fairweather[1], Philip A. Bland[1], Hadrien A. R. Devillepoix[1], Eleanor K. Sansom[1], Martin Cupak[1], Patrick M. Shober[1], Gretchen K. Benedix[1]

[1]Space Science and Technology Centre, School of Earth and Planetary Science, Curtin University, 314 Wark Ave, Bentley, WA, Australia 6102
[*]seamus.anderson@postgrad.curtin.edu.au

**Abstract**

We report the first-time recovery of a fresh meteorite fall using a drone and a machine learning algorithm. A fireball on the 1st April 2021 was observed over Western Australia by the Desert Fireball Network, for which a fall area was calculated for the predicted surviving mass. A search team arrived on site and surveyed 5.1 km$^2$ area over a 4-day period. A convolutional neural network, trained on previously-recovered meteorites with fusion crusts, processed the images on our field computer after each flight. Meteorite candidates identified by the algorithm were sorted by team members using two user interfaces to eliminate false positives. Surviving candidates were revisited with a smaller drone, and imaged in higher resolution, before being eliminated or finally being visited in-person. The 70 g meteorite was recovered within 50 m of the calculated fall line using, demonstrating the effectiveness of this methodology which will facilitate the efficient collection of many more observed meteorite falls.

## 1. Introduction

Besides recording the conditions of the proto-planetary nebula and the early Solar System, meteorites also offer insights into the contemporary physical and chemical compositions of Asteroids and other terrestrial bodies (Cuzzi et al. 2008; Nakamura et al. 2011). Some of these meteorites fall in regions on Earth where fireball observatory networks are active, making it possible to record the trajectory of the fireball as it ablates material from the originating meteoroid. For some fireballs, this data can then be used to simulate both forward and backward in time to predict where the resulting meteorite landed on Earth and where the meteoroid originated in the solar system. Thus, recovering and analyzing these 'orbital meteorites' with constrained, prior orbits provides an incredibly unique insight into the geology of the asteroid belt and the nature of mass transfer between the belt and the inner solar system. The Desert Fireball Network (DFN) (Bland et al. 2012; Howie et al. 2017) is one of many organizations (Oberst et al. 1998; Spurný et al. 2006; Trigo-Rodríguez et al. 2006; Olech et al. 2006; Colas et al. 2015; Devillepoix et al. 2020) that makes this possible.

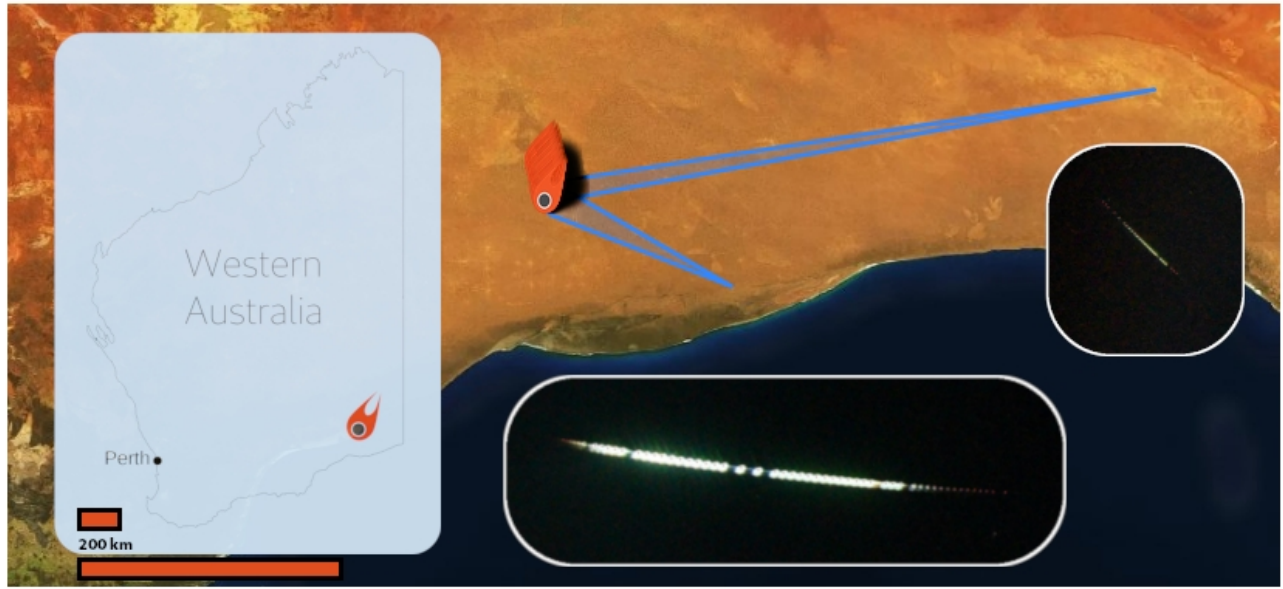
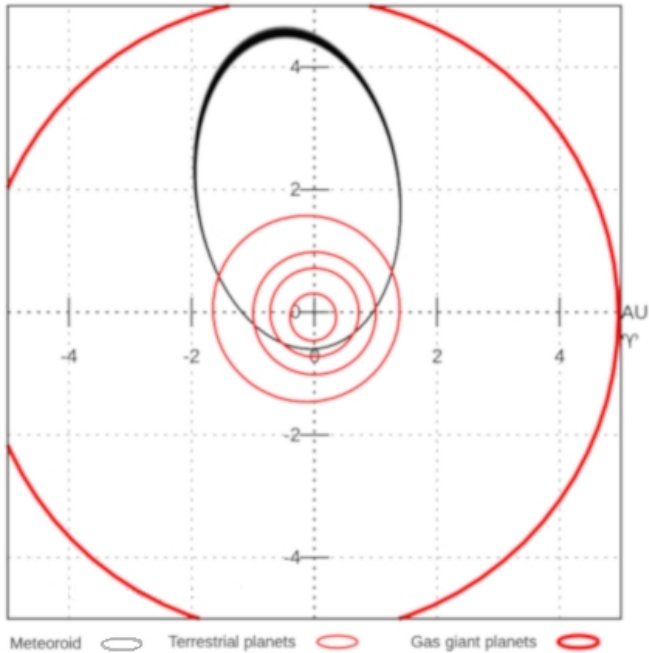
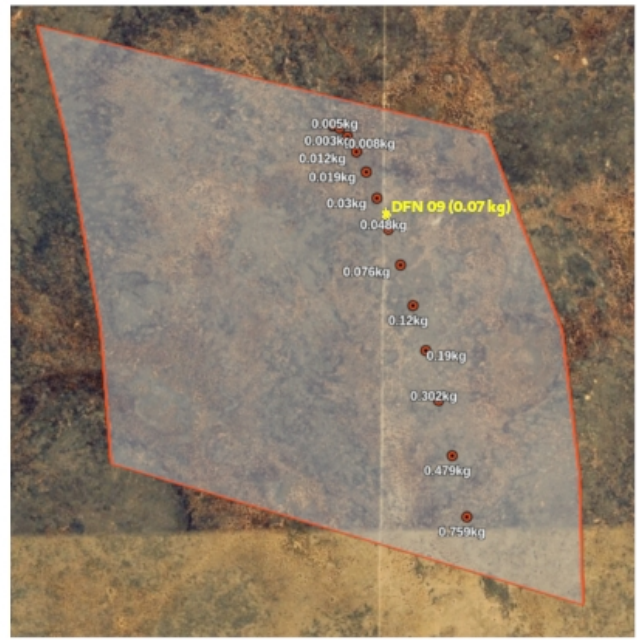

Figure 1. The DFN 09 meteorite fall at Kybo Station, Western Australia. (Clockwise from top) Fireball observations from DFN camera stations at Mundrabilla Station and O'Malley Siding, and their location within WA; The 90% certainty searching area (transparent white), the best fit fall line (red markers), and the location of the recovered meteorite (yellow star); Pre-impact orbit for the DFN 09 meteoroid.

In the past, recovering meteorites within a predicted area often consisted of 4-6 people walking 5-10 m apart, sweeping the area until the meteorite is found. This was labor intensive and suffered from a relatively low success rate of ~20%, since an entire fall zone could rarely be covered in one trip. This has prompted efforts by multiple groups to recover meteorites using drones and machine learning which massively reduces the time and labor required (Citron et al. 2017; 2021; Zender et al. 2018; AlOwais et al. 2019). Our previous efforts (Anderson et al. 2019; 2020) have shown the ability of our methodology to positively identify already-recovered fresh meteorites, awaiting an opportunity to fully test our searching strategy.

On the night of April 1$^{st}$ 2021 such an opportunity presented itself as a meteorite fell over the Western Nullarbor on the Lintos Paddock of Kybo Station, Western Australia (Figure 1). Unfortunately, only two DFN camera stations, both to the East, were able to capture the event which caused the relatively short fall line to have considerable longitudinal uncertainty, expanding the area with a 90% likelihood of containing the meteorite to a total of 5.1 km$^2$ (Figure 1), with a predicted mass of the meteorite was between 150 and 700 g. These fall conditions were promising enough to warrant a fieldtrip to survey the entire fall zone with a drone. The first three days we spent onsite consisted of surveying with a drone, and processing data with our machine learning algorithm. On the fourth and final day we visited meteorite candidates with the drone and in person, and recovered the 70 g meteorite.

## 2. Methods

### 2.1 Fireball Observations and Modelling

Although the detailed analysis of the fireball and its orbit will be the subject of a future publication, here we briefly explain the data and methods that allowed us to predict the searching area. These analyses were done prior to the recovery, and have not been revised since. On 1 April 2021, two Desert Fireball Network observatories imaged a bright 3.1 s fireball, which was automatically reported to the DFN team by our automated detection software (Towner et al. 2020), while astrometric calibration is performed following the methodology detailed in Devillepoix et al. (2018).

In total, 78 data points were recorded from just two observatories located at Mundrabilla station and O'Malley siding, 149 km and 471 km from the end point, respectively (Figure 1). The nominal trajectory started at 87 km altitude at 25.4 km/s, and the bolide was observed down to 25 km at 8.4 km/s, on a 64 deg slope. Because of the distance of the viewpoints and a low convergence angle of planes of 28 deg, observation conditions were not ideal, resulting in a poorly constrained trajectory. To estimate the variability in the trajectory, in a similar manner to Devillepoix et al. (2021), we use a Monte Carlo approach, randomising astrometric observations within errors, generating 1000 clones. This analysis shows a 500 m standard deviation on the end point.

For estimating the surviving mass of the object, we use the alpha-beta method from Gritsevich et al. (2012) and Sansom et al. (2019). The noisy velocity data resulted in a poorly constrained surviving mass between 150 and 700 g, assuming a spherical to rounded brick shape, a bulk density of 3.5 g/mL, and a shape change parameter of 2/3. We model the atmospheric conditions numerically using the Weather Research and Forecasting (WRF) model version 4.0 with the Advanced Research WRF dynamic solver (Skamarock et al. 2019) with 3 model runs starting on 2021-04-01 at 00UT, 06UT, and 12UT, all giving similar models.

Using these atmosphere models, we then propagate bright flight observations to the ground using the dark flight model from Towner et al. (2021). The uncertainty on the mass, and the uncertainty on the positions are the dominant sources of uncertainty. Thanks to the dominant wind being more or less in the same direction as the fireball, the small mass end of the fall line was somewhat compacted. From these simulations, we defined two search areas at different confidence levels: a 90 % confidence level yielding a search area of 5 km² (Figure 1), and a 99 % confidence level which gives a significantly larger search area of 8 km².

### 2.2 Drone Surveying and Machine Learning

Our field-based work is a continuation of the methodology presented in Anderson et al. (2020). For this trip we used a DJI M300 drone with a Zenmuse P1 camera (44 MP) to survey the 5.1 km² fall line at 1.8 mm/pixel with 20% overlap among images in each direction, which took between 2.5 and 3 days to complete. This ground sampling distance would image the meteorite at an apparent size between 20-65 pixels in diameter, due to inherent uncertainties in meteoroid properties during the fall. We processed the data on-site using a desktop computer with an RTX 2080 Ti GPU. Our algorithm could process one flight's worth of images (~30 min) in approximately 65 min.

Table 1. The model architechture for our convolutional neural network (CNN). ReLU means Rectified Linear Unit, a common activation function for CNNs

| Layer | no. filters/ nodes | kernel | stride | Activation Function |
|---|---|---|---|---|
| Convolutional 2D | 30 | 3x3 | 1 | ReLU |
| Batch Normalization | | | | |
| Max Pooling | | 2x2 | 2 | |
| Convolutional 2D | 60 | 3x3 | 1 | ReLU |
| Batch Normalization | | | | |
| Max Pooling | | 2x2 | 2 | |
| Convolutional 2D | 120 | 3x3 | 1 | ReLU |
| Batch Normalization | | | | |
| Max Pooling | | 2x2 | 2 | |
| Convolutional 2D | 240 | 3x3 | 1 | ReLU |
| Batch Normalization | | | | |
| Max Pooling | | 2x2 | 2 | |
| Flatten | | | | |
| Dense | 1000 | | | ReLU |
| Dropout | 0.5 | | | |
| Dense | 150 | | | ReLU |
| Dropout | 0.5 | | | |
| Dense | 1 | | | Sigmoid |

Our algorithm works by taking a full 44 MP image and splitting it into tiles (125 x 125 pixels) with a 70 pixel overlap in each direction, to ensure the meteorite has a chance to appear fully in at least one tile. Each tile is fed into a binary image classifier constructed using python and keras (Chollet et al. 2015) (architecture shown in Table 1) which scores each tile from 0 (non-meteorite) to 1 (meteorite). To obtain True (meteorite) training data, we relied on our library of meteorite images from previous trips as well as taking new images of real meteorites (Camel Donga and Wiluna, supplied by the Western Australian Museum) at the fall zone. We repeated meteorite image collection each day and for each weather condition that affected illumination. Our True pool was comprised of all the meteorites we imaged on-site and an equal number again sourced from our library, totalling ~100,000 tiles. To create False tiles (non-meteorites) we took images from the surveyed fall zone at random, checked to ensure there were no meteorites in the frame, and split them into tiles. These totalled to more than 1 M depending on the day of the flights being sampled.

To maintain a balanced training set, while also sampling as much of our dataset as possible, we employed what we call 'rotation training'. To form a training set, we use a constant 80% of our True pool (with 20% kept for validation) and randomly select an equal number from the False pool, then we train for 5 epochs (rounds of training). After this we deselect these False tiles and randomly select a new False set from the pool, then repeat the training process until the model had worked through the False pool twice. The validation set contained only True tiles, so that we could monitor the most important performance metric: meteorite detection chance. We initially trained the model until it rotated through the False pool twice. By the end of training, we achieved a training accuracy of 99.93%, and a validation accuracy (meteorite detection chance) of 91%.

As we predicted on each flight from our first day of surveying, we monitored the quality of the prediction by viewing the distribution of confidence values for a given image, in the form of a histogram (Figure 2). When the distribution for an image resembled that of Figure 2A, we were satisfied, while images that produced a histogram like Figure 2B were flagged as 'problem images'. We believe they contain physical features from the survey area that were not yet included in the training data, causing the model to infer generously, creating more unnecessary false positives to sort later in the process. We inspected some of these images, split them into tiles and added them to the False pool. We later retrained on this augmented False pool, though only for two epochs and one rotation. Using the retrained model we would re-predict on these problem images, which usually resulted in a more palatable confidence distribution.

We inspected meteorite candidate tiles in 4 stages (Figure 3). The first stages were identified by the model to have a confidence >0.7, and we inspected them using the 3x3 grid graphical user interface (GUI) described in Anderson et al. (2020), which was ideal for eliminating obvious false positives. The user would be given a 3x3 grid displaying 9 tiles, with some being sourced from the True pool as tests for the user, to characterize their own performance. They identified interesting tiles by typing the corresponding key on the number pad before moving to the next set. Uninteresting tiles were added to the false pool, while those of interest became second stages that were further inspected in a separate image viewer that allowed pan and zoom control over the whole image. If the candidate was still of interest, it became a third stage. To inspect the third stages, we compiled all of their GPS coordinates from one survey flight and planned a new waypoint flight using our DJI Mavic Pro drone and GS Pro app. The Mavic was ideal for this task as it could lower to ~ 1 m altitude with much lower assumed risk than the more expensive M300. When the autopilot flew the drone to a waypoint (at ~20 m altitude), we paused the mission and manually lowered the drone directly above the candidate, using the original survey image and prediction box as a guide. If the entire team could confidently eliminate that candidate we would remove it, otherwise we took 2-3 pictures for later inspection and proceeded via autopilot to the next waypoint. Since live-feed transmission from the drone provided a lower resolution than its camera, we reviewed the images on our field computer in camp. Any candidates that survived this scrutiny proceeded to the fourth stage: in-person inspection.

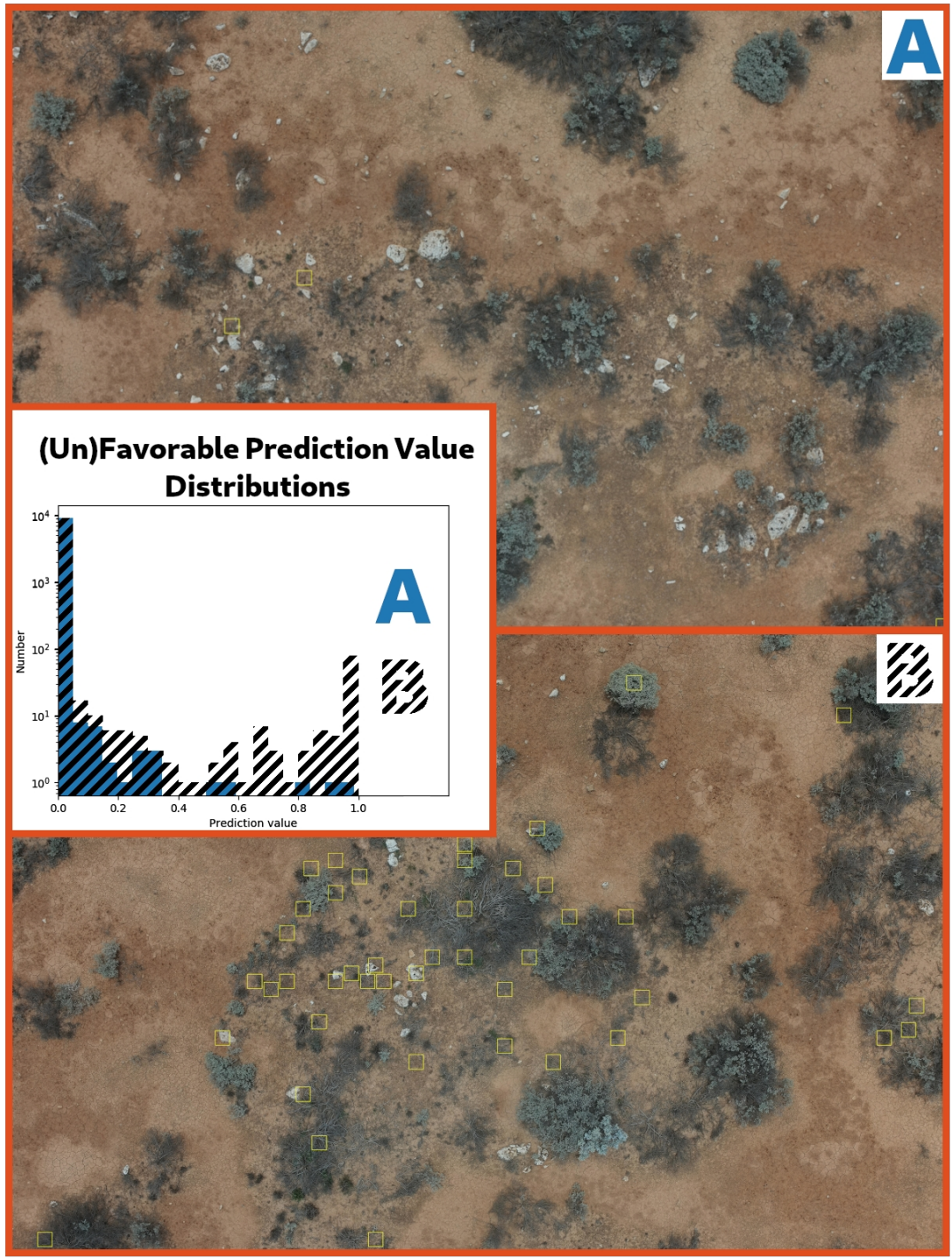

Figure 2. Favorable and unfavorable prediction distributions from two images. Given a 70% confidence threshold, Image/Distribution A will return 3 meteorite candidates, while Distribution B will return >100 candidates. B-like images are later used for retraining. For clarity, the number of detections displayed in image B is capped at 50.

## 3. Results and Discussion

Using the above methodology, we recovered the meteorite <50 m from the best-fit fall line, appearing in the 88th image from the 3rd flight on the 1st day, with an apparent diameter of 27 pixels and a confidence score of 1: a perfect match. Although we have not yet classified the meteorite, its fusion crust (Figure 4) resembles that of other chondrites. It consists of one 70 g piece, approximately 5x4x3 cm, with a preferentially smoothed side.

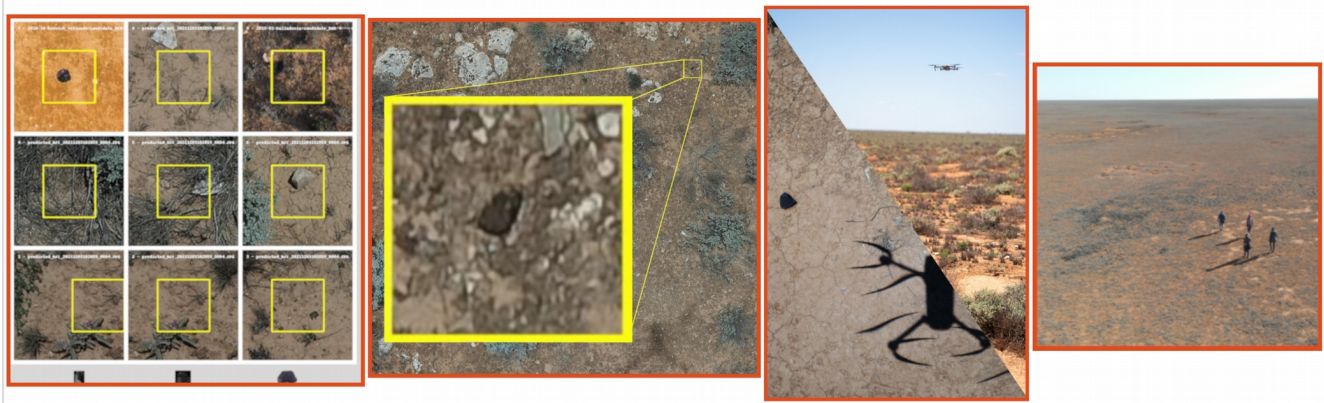

Figure 3. The four stage process for eliminating false positives and verifying meteorite candidates. (From Left to Right) 1) Grid GUI. 2) Zoom-pan GUI. 3) Drone visit. 4) In-person visit.

While this meteorite is a great discovery that will hopefully spur the efficient collection of further meteorites, we also made some important discoveries about or methodology. Our data processing rate for both the machine learning algorithm and manual candidate sorting must be improved in the future, or the length of the trips must be extended as the meteorite appeared on the 3rd/43 survey flights. We were able to process 4 flights, totalling 5096 images, which produced 46,501,000 tiles for our algorithm, identifying 56384 first stage candidates (Figure 3). Using our candidate elimination process, we produced 259 second stage candidates, visiting 38 of these as third stages with the Mavic Pro, finally searching for 4 candidates in person (fourth stage). If we were instead required to process all 57,255 images, we would not have been able to complete it onsite, and would have to return on a later trip to follow up on meteorite candidates. That being said, the meteorite was located <50 m from our 'ideal' fall line, which is surprising considering the side-to-side uncertainty in the fall line. With this in mind, we may in the future prioritize searching the area immediately around the ideal fall line.

Although we recovered the meteorite, we did not really train a meteorite detection algorithm. Instead, what we really created an anomaly detector, in this case trained for the Nullarbor. During the course of devising this strategy, we have encountered false positives such as tin cans, bottles, snakes, kangaroos, and piles of bones from multiple animals. We also notice that when we predict on survey images taken directly over our campsite, the algorithm ferociously identifies our items and equipment, none of which is represented in the training data. We hope that our findings and methodology prove useful for training neural networks in other low-occurrence or anomaly detection problems, such as wildlife monitoring, or search and rescue.

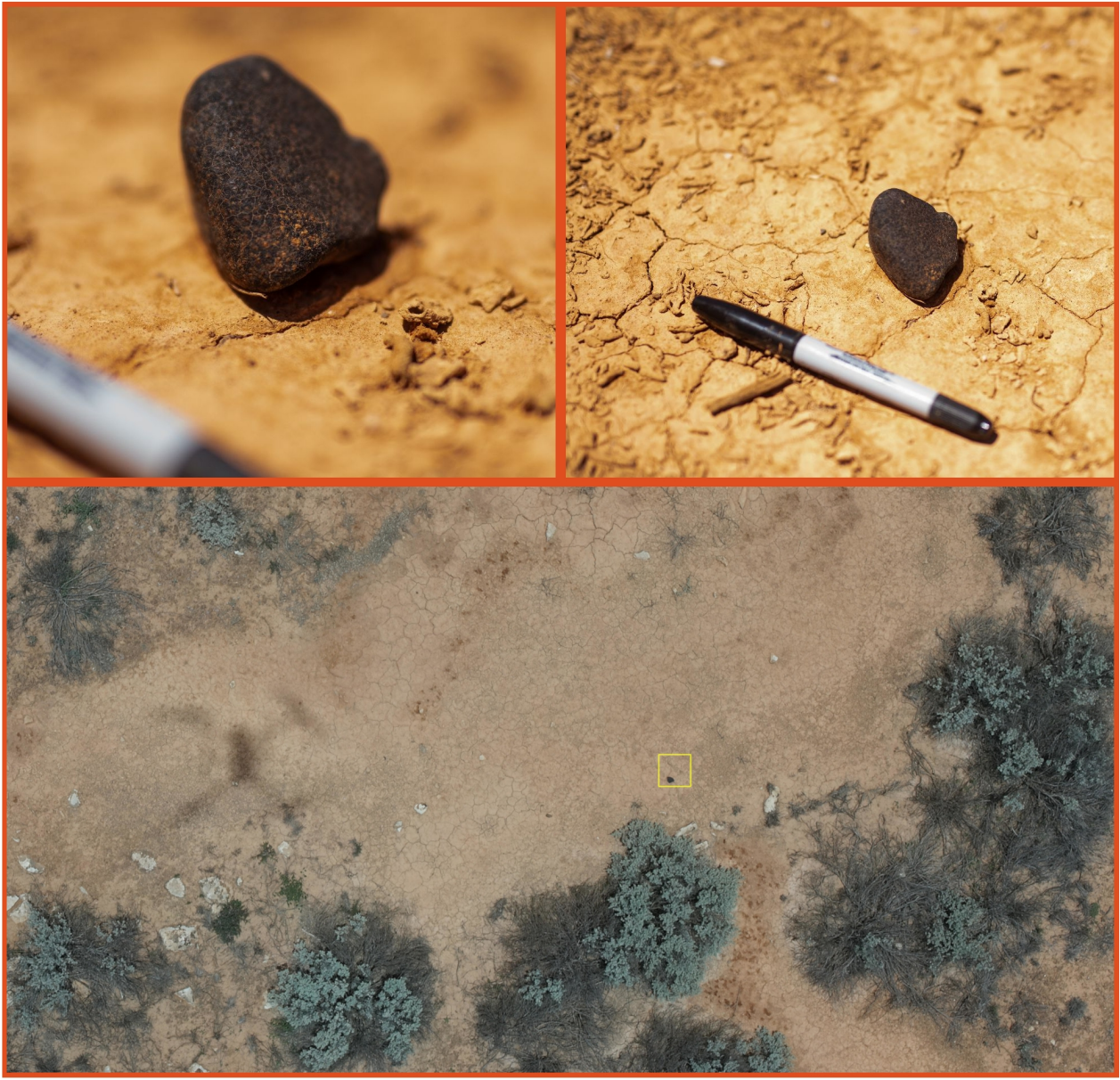

Figure 4. The recovered meteorite as seen in person (top two), and from the survey drone (bottom one). For scale, a 15 cm long felt pen is placed next to the meteorite (top right). The yellow box in the bottom image is 22 cm on one side.

**Acknowledgements**

This work was funded by the Australian Research Council as part of the Australian Discovery Project scheme (DP170102529, DP200102073), and receives institutional support from Curtin University. Fireball data reduction is supported by resources provided by the Pawsey Supercomputing Centre with funding from the Australian Government and the Government of Western Australia. The DFN data reduction pipeline makes intensive use of Astropy, a community-developed core Python package for Astronomy (Price-Whelan et al. 2018). We would like to thank Geoff Deacon and Peter Downes at the Western Australian Museum for loaning us meteorites from the collection.